\documentclass[prl,aps,twocolumn,showpacs]{revtex4}
\usepackage{amsmath} \usepackage{graphicx}

\begin{document}

\title{Entangling strings of neutral atoms in 1D atomic pipeline
  structures}

\author{U.~Dorner$^1$, P.~Fedichev$^1$, D.~Jaksch$^1$,
  M.~Lewenstein$^2$, and P.~Zoller$^{1,2}$}

\affiliation{${}^1$Institute for Theoretical Physics, University of
  Innsbruck, A--6020 Innsbruck, Austria}

\affiliation{${}^2$Institut f{\"u}r Theoretische Physik,
  Universit{\"a}t Hannover, D-30167 Hannover, Germany}

\begin{abstract}
  We study a string of neutral atoms with nearest neighbor interaction
  in a 1D beam splitter configuration, where the longitudinal motion
  is controlled by a moving optical lattice potential. The dynamics of
  the atoms crossing the beam splitter maps to a 1D spin model with
  controllable time dependent parameters, which allows the creation of
  maximally entangled states of atoms by crossing a quantum phase
  transition.  Furthermore, we show that this system realizes
  protected quantum memory, and we discuss the implementation of one-
  and two-qubit gates in this setup.
\end{abstract}


\maketitle

The recent development of optical and magnetic microtraps allows the
confinement of cold atoms in effective 1D ``pipeline'' geometries,
where the transverse quantum motion is frozen out \cite{1D}. Variants
of these trap designs promise the realization of beam splitters, and
thus atomic interferometry ``on a chip''. Usually one envisions that
atoms are injected one by one into these pipelines, where the source
of cold atoms is provided by a Bose Einstein condensate. Instead we
will study below collective beam splitter setups which allows the
\textit{generation of entangled strings of atoms} in 1D trapping
configurations with applications in interferometry and quantum
computing.

\begin{figure}[tbp]
  \centering \includegraphics[]{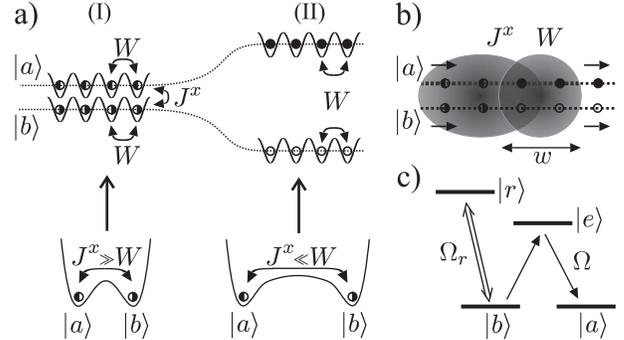} \caption{\textit{External beam
      splitter}: a) Atoms before (I) and after (II) the separation.
    The nearest neighbor interaction is denoted by $W$ and $J^x$ is
    the hopping matrix element between the two states $|a\rangle$ and
    $|b\rangle$ of the transverse trapping potential.
    \textit{Internal beam splitter}: b) atoms in two different
    internal states $|a\rangle$ and $|b\rangle$ enter the beam
    splitter. The internal states $|a\rangle$ and $|b\rangle$ are
    coupled by a Raman transition [see c)] with a Rabi frequency
    $J^x=\Omega$. A laser excited Rydberg state $|r\rangle$ realizes
    the offsite interaction $W$ with $w$ denoting the width of the
    interaction zone.}
\label{fig1}
\end{figure}

To this end, we assume that the longitudinal motion of the atoms is
controlled by storing atoms in a 1D optical lattice potential
generated by a standing light laser field.  In the transverse
direction the particles are confined by a a double well potential (see
Fig.~\ref{fig1}a) where we assume that the optical lattice stores
exactly one atom per lattice site (i.e. one atom per double well).
The preparation of such a Mott insulating state has been reported in a
recent experiment, by loading of atoms from a Bose Einstein condensate
via a superfluid - Mott insulator quantum phase transition
\cite{MottExp}.  This setup by itself is an interesting extension of
the standard ``interferometry on a chip'', as it eliminates
collisional shifts since atoms stored on different lattice sites never
collide.  Furthermore, the atoms are supposed to be initially in the
ground state which is a spatial superposition of the particles in the
two transverse wells (region (I) of Fig.~\ref{fig1}a).  By moving the
lattice we can drag the atomic chain ``by hand'' across the beam
splitter while we increase the distance between the transverse wells
adiabatically depending on the position of the atoms (i.e. we decrease
the tunneling $J^x$ between the wells, see region (II) of
Fig.~\ref{fig1}a).  On the other hand, the use of optical lattices
allows the engineering of coherent interactions between adjacent atoms
(nearest neighbor interaction $W$ in Fig.~\ref{fig1}a). This can be
obtained either by cold collisions and moving optical lattices
\cite{CCC,Molmer,Mandel}, the remarkably strong dipole-dipole
couplings of laser excited Rydberg atoms \cite{Rydberg}, or by
dipole--dipole coupling of cold heteronuclear molecules
\cite{DeMille}.  Together with appropriate detection methods like
fluorescence imaging these controllable interactions provide us with
the tools to generate entanglement of the 1D chain of atoms.

We will study the dynamics of the beam splitter setup indicated in
Fig.~\ref{fig1}. In particular, we will investigate (i) how to
generate a maximally entangled state of atoms. (ii) We will establish
the formal equivalence of our model with well-studied models of spin
chains.  In particular, we will show that the system dynamics is a
physical realization of a textbook model of a quantum phase transition
with completely controllable (time dependent) parameters
\cite{Sachdev}. Thus our setup provides an example of engineering a
maximally entangled state from a product state via a quantum phase
transition. (iii) Finally, the present setup implements the spin
analogue \cite{Levitov} of Kitaev's protected quantum memory
\cite{Kitaev}, where qubits are represented by Majorana fermions,
which provide a stable way to store quantum information due to an
excitation gap \cite{Remark}.  Our setup allows to perform single and
(collectively enhanced) two qubit operations.

We consider a 1D chain of $N$ atoms with modes $|a\rangle $ and
$|b\rangle $ stored in an optical lattice with a lattice constant
$\lambda /2$ determined by the wave length $\lambda $ of the laser.
The modes correspond either to two spatial modes in a double well
structure, where the tunnelling provides a coupling (external beam
splitter in Fig.~\ref{fig1}a), or to two internal atomic states
connected via a Raman process (cf.~Fig.~\ref{fig1}b,c). We suppress
hopping of the atoms between adjacent lattice sites by a sufficiently
large potential barrier. This leads to an onsite interaction
$U\rightarrow \infty $ and we assume to have commensurate filling of
one particle per lattice site. Following \cite{CCC,MottIns} we derive
a Hubbard Hamiltonian
\begin{eqnarray}
&&H(t) =  2\sum_{l=1}^{N-1}W_{l}(t)\left( a_{l+1}^{\dagger
}a_{l+1}a_{l}^{\dagger }a_{l}+b_{l+1}^{\dagger }b_{l+1}b_{l}^{\dagger
}b_{l}\right) \nonumber \\
&&-\sum_{l=1}^{N}\left( J_{l}^{x}(t)\left( a_{l}^{\dagger
}b_{l}+a_{l}b_{l}^{\dagger }\right) +J_{l}^{z}(t)\left( a_{l}^{\dagger
}a_{l}-b_{l}^{\dagger }b_{l}\right) \right). 
 \label{Hamil}
\end{eqnarray}%
Here $J_{l}^{x}$ describes coupling between $|a\rangle $ and
$|b\rangle $ while the operators $a_{l}$, $b_{l}$ are bosonic
annihilation operators for particles in these modes at site $l$ with
$[a_l,b_j]=[a_l,b_j^\dagger]=0$. A term $J_{l}^{z}$ emerges from an
additional state dependent superimposed trapping potential. We
introduce the spin notation $\sigma _{l}^{x}=a_{l}^{\dagger
}b_{l}+a_{l}b_{l}^{\dagger }$, $\sigma _{l}^{z}=a_{l}^{\dagger
}a_{l}-b_{l}^{\dagger }b_{l}$ and $\sigma
_{l}^{y}=i(a_{l}b_{l}^{\dagger }-a_{l}^{\dagger }b_{l})$ which for
$n_{l}=a_{l}^{\dagger }a_{l}+b_{l}^{\dagger }b_{l}\equiv 1$ are Pauli
operators and rewrite the Hamiltonian (\ref{Hamil}) as
$H_S(t)=\sum_{l=1}^{N\!-\!1}W_{l}(t)\sigma _{l}^{z}\sigma
_{l\!+\!1}^{z}\!-\!\sum_{l=1}^{N}\mathbf{J}_{l}(t)\cdot
\mathbf{\sigma}_{l}.$ Thus our setup is formally equivalent to an
Ising chain of $N$ spins in a magnetic field
$\mathbf{J}_{l}=(J_{l}^{x},J_{l}^{y},J_{l}^{z})$ \cite{Duan,Remark2}.

\textit{Entanglement via quantum phase transition:} Moving a string of
atoms from left to right in the setup of Fig.~\ref{fig1}a, or
switching the lasers in Fig.~\ref{fig1}b,c amounts to a time dependent
change of the parameters from the
large tunneling limit $J_{l }^{x}(t=0)\gg |W_l|$ to small tunnelling $%
J_{l }^{x}(t=T)\rightarrow 0$. In the following we assume that
$J_{l}^{y,z}=0$ except it is stated differently. In
the homogeneous case (i.e. $J^x_l = J^{x}$, $W_l = W$) the variation
of $J^x$ amounts to crossing the critical point at $J^{x}=W$ of a
quantum phase transition \cite{Sachdev}. Assuming that the atoms are
initially prepared in the product state $|\uparrow \uparrow \cdots
\uparrow \rangle ^{x}$ with $|\uparrow \rangle _{l}^{x}\sim |a\rangle
_{l}+|b\rangle _{l}$ a superposition state of the two modes which is
for $W=0$ the (paramagnetic) ground state of $H_S$.  Under adiabatic
variation of parameters the system will remain in the ground state and
evolve according to ($W<0$)
\begin{eqnarray}
|+\rangle \!\equiv |\uparrow \uparrow \cdots \uparrow \rangle ^{x}
&\longrightarrow &(|\uparrow \uparrow \cdots \uparrow \rangle
^{z}+|\downarrow \downarrow \cdots \downarrow \rangle
^{z})/\sqrt{2}  \nonumber \\
&\equiv &\left( |0\rangle +|1\rangle \right)/\sqrt{2} ,
\label{trafo1}
\end{eqnarray}%
where the states $|\uparrow \rangle _{l}^{z}=|a\rangle _{l}$ and $%
|\downarrow \rangle _{l}^{z}=|b\rangle _{l}$ correspond to the atoms
being in the upper or lower branch of the beam splitter of Fig.~\ref{fig1}. The states $%
|0\rangle $ and $|1\rangle $ are the two degenerate (ferromagnetic)
ground states of the $H_S$ for $J^x=0$ with all atoms in either one or
the other arm of the beam splitter (see Fig.~\ref{fig1}). Thus the
initial \emph{product state }is transformed to a \emph{maximally
  entangled state} via a quantum phase transition.  The intuitive
physical picture behind (\ref{trafo1}) is as follows.  Consider atoms
moving across the beam splitter one by one.  The first atom of the
string will end up in the state $|\uparrow \rangle ^{z}+|\downarrow
\rangle ^{z}$, and attract the second atom.  This leads to a state of
the form $|\uparrow \uparrow \rangle ^{z}+|\downarrow \downarrow
\rangle ^{z}$. After the last atom has left the interaction zone the
maximally entangled state $|0\rangle +|1\rangle $ has been created.

In the following we discuss the validity of the adiabatic
approximation (Eq.~(\ref{trafo1})) and thus the usefulness of this
scheme by studying the scaling of the fidelity $F=|\langle \psi
_{\text{id}}|\psi (T)\rangle|^{2}$ as a function of the length of the
string $N$ and the time variation of $J_l^x(t)$ and $W_l(t)$.  Here
$F$ compares the state $|\psi (T)\rangle $ obtained from a time
dependent integration of the Schr\"{o}dinger equation with the ideal
state $|\psi _{\text{id}}\rangle \sim |0\rangle +|1\rangle $. This
will be first done numerically, followed by analytical calculations
and estimates.

\begin{figure}[tbp]
  \centering \includegraphics[]{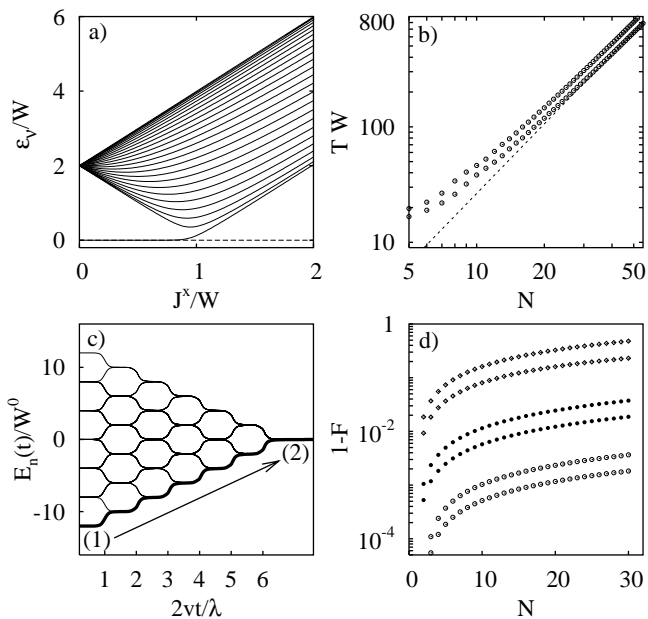} \caption{\textit{Homogeneous
      setup $J_l^x \equiv J^x$, $W_l \equiv W$}: a) Elementary
    excitations $\protect\varepsilon_\protect\protect\nu$ for $N=25$
    against $J^x/W$. The dashed horizontal line indicates the ground
    state energy. b) Upper bound $F_2$ and lower bound $F_1$ for the
    time $T$ yielding a fidelity of $F=95\%$ as a function of $N$. The
    dashed line illustrates the $N^2$ scaling predicted analytically.
    \textit{Inhomogeneous setup $J_l^x$, $W_l$}: c) Instantaneous
    eigenenergies $E_n$ for $N=6$. d) Upper and lower bounds for the
    infidelity $1-F$ against $N$ for constant sweeping speed $v=0.01
    \protect\lambda W$ and different interaction zone widths $w=0.1
    \protect\lambda$ (diamonds), $w=0.2 \protect\lambda$ (closed
    circles) and $w=0.4 \protect\lambda$ (open circles) and similar
    for $J^x$.} \label{fig2}
\end{figure}

Before entering the time dependent case, we note that for the time
independent case the Hamiltonian $H$ has been studied extensively
\cite{Sachdev,Pfeuty}. For $J^z_l=0$ it can be fermionized and one
obtains $H_F=\sum_{\nu }\varepsilon _{\nu }(f_{\nu }^{\dagger }f_{\nu
}-1/2)$ with the elementary excitation energies $\varepsilon _{\nu }$
and fermionic annihilation (creation) operators $f_{\nu }$ ($f_{\nu
}^{\dagger }$). The spectrum for the homogeneous case is shown in
Fig.~\ref{fig2}a. For large $N$ the spectrum of the elementary
excitations is characterized by a gap $\Delta =2|W-J^{x}|$ for the
energetically low lying quasi particles with the exception (arising
from the free end boundary conditions) that the first excited state
becomes degenerate with the ground or vacuum state (here, we do not
take into account the second term in $H_F$, i.e. the vacuum state has
zero energy) for $|W| \gg J^{x}$ (cf.~Fig.~\ref{fig2}a). For $J^{x}=0$
the two cat type ground states $|0\rangle +|1\rangle$ and $|0\rangle
-|1\rangle$ correspond to the vacuum and the first excited state of
the fermionized system, respectively.

In Fig.~\ref{fig2}b we plot the numerically calculated operation time
$T$ required to perform (\ref{trafo1}) with a fidelity of $F=95\%$ for
linearly changing the homogeneous couplings $J^{x}(t)$ against $N$
($W=$ const.). For $N>20$ we find a (polynomial) scaling of $T$ for a
given infidelity $1-F\sim N^{2}$ (cf.~Fig.~\ref{fig2}b) in agreement
with the analytical results below. By optimizing the time dependence
of $J^{x}(t)$ we can speed up by the entanglement process
significantly.

A discussion of the spatially inhomogeneous situation where $J^x_l$
and $W_l$ vary as a function of $l$ corresponding closer to the setup
of Fig.~\ref{fig1} is given in Figs.~\ref{fig2}c,d.  For increasing
time the string is moved across a zone of non vanishing $W_{l}$ with a
maximum $W^{0}$ and a width $w$.  Simultaneously, $J_{l}^{x}(t)$, is
decreased from the initial value to $J_{l}^{x}(T)\approx 0$ over a
comparable ``width'' as $w$ for all sites. The corresponding
instantaneous time dependent energy levels are shown in
Fig.~\ref{fig2}c. Following the lowest energy curve in this diagram
adiabatically from (1) to (2) corresponds to $|+\rangle \rightarrow
|0\rangle +|1\rangle$.  Fig.~\ref{fig2}d shows the infidelity $1-F$
for finite sweeping speed $v$ against $N$ for different widths $w$ of
the interaction zone. The infidelity $1-F$ decreases rapidly with
increasing $w$ and scales exponentially with $N$ for $w\ll \lambda N$.
For $w \gtrsim N\lambda /2$ the above scaling $1-F\sim N^{2}$ is
restored.

The numerical calculations behind Fig.~\ref{fig2} were based on a
\textit{time independent} Jordan-Wigner transformation of $H_S$,
yielding a quadratic Hamiltonian in fermionic destruction (creation)
operators $\gamma_{\nu }$ ($\gamma_{\nu }^{\dagger}$). By introducing
Majorana operators $c_{2\nu -1}=(\gamma _{\nu }+\gamma _{\nu
}^{\dagger })/2,\,c_{2\nu }=(\gamma _{\nu }-\gamma _{\nu }^{\dagger
})/(2i)$ \cite{Kitaev} we obtain
$H_{t}=i\mathbf{c}^{T}\mathbf{A(t)}\mathbf{c} $, where the components
of $\mathbf{c}$ are the Majorana operators and $\mathbf{A(t)}$ is a
$2N\times 2N$ real antisymmetric tridiagonal matrix.  The
\textit{linear} Heisenberg equations of motion, $\dot{\mathbf{c}}=
\mathbf{A(t)}\mathbf{c}$, are then solved numerically. We note that
diagonalizing $\mathbf{A}$ in the time independent case yields $H_F$.
For the fidelity $F$ we use an approximate expression which can be
derived as follows: The state $|0\rangle +|1\rangle$ is the vacuum
state of the fermionized system at $t=T$. The completeness relation
yields $F(T)=1-\sum_{\mathbf{n}\neq \mathbf{0}}|\langle
\mathbf{n}|\psi (T)\rangle |^{2}$. Here $|\mathbf{n}\rangle
=|n_{1},\ldots ,n_{N}\rangle $ with $n_{\nu }=0,1$ the occupation
numbers of the instantaneous eigenstates of $H_{T}$ corresponding to
an energy $\epsilon_{\nu}$. The sum in this
expression can be reordered, and we obtain $F(T)=1-\sum_{m=1}^{N}P(m)$, where $%
P(m)=\sum_{\mathbf{n}}|\langle \mathbf{n}|\psi (T)\rangle |^{2}\delta
_{m,\sum n_{i}}$ is the probability of having $m$ elementary
excitations in the system at time $T$. By solving the above equation for $%
\mathbf{c}$ we can in principle calculate the quantities $A_{l}\equiv
\langle (\sum_{m=1}^{N}f_{m}^{\dagger }f_{m})^{l}\rangle
=\sum_{m=1}^{N}P(m)m^{l}$. The fidelity $F$ is then given by the
solution of a system of $N$ linear equations. An approximate fidelity
$F_{l}$ can be obtained by neglecting the probabilities $P(k)$ with
$k>l$. We restrict ourselves to $l=1,2$ and find $F_{1}=1-A_{1}$ and
$F_{2}=1-(3 A_{1}-A_{2})/2$. The exact fidelity is bounded by these
quantities: $F_{1}\leq F\leq F_{2}$. Compared to a calculation in the
spin picture which requires the solution of $\sim 2^{N}$ equations the
calculation of $A_{1}$ and $A_{2}$ can be done by solving a system of
$\sim N^{2}$ differential equations.

Let us turn to the more technical point of analytically estimating the
scaling of the fidelity $F$ when the phase transition point is crossed
by linearly changing $J^{x}=\Theta t+W$ with $\Theta =$const. First we
note that there are no transitions between the ground and the first
excited state since they have opposite parity. Close to the phase
transition point the energy gap to the remaining excitations $\Delta
\approx 0$ and therefore at the time $t=-t_{\ast }$ the evolution of
the system ceases to be adiabatic and excitations start to be
populated. The adiabaticity is restored again at the time $t\sim
t_{\ast }$, when the gap $\Delta $ becomes sufficiently large to
prevent further excitations. Then, the relaxation of the new phase
occurs separately within different domains, whose sizes are given by
the value $l_{0}(t_{\ast })$ of the correlation length at the time
$t_{\ast }$. Close to the phase transition $l_{0}\sim \Delta ^{-1/2}$
and therefore the domain size scales like $l_{0}(t_{\ast })\sim \Theta
^{-1/2}$. The quench through the phase transition point can only be
adiabatic if the
characteristic size of the domain formed exceeds the size of the system $%
L\sim N$ and therefore $l_{0}(t_{\ast })\agt L$, which gives the
scaling condition $\Theta \alt W^{2}/N^{2}$, or $WT\sim N^2$.

\textit{Quantum computing model with protected quantum memory:} In the
case $W<0$ the ferromagnetic superposition state is very sensitive to
homogeneous distortions of the form $J_{l}^{z}=J^{z}$ which induce a
relative phase shift $\exp (i2N\int_{0}^{\tau }dtJ^{z}(t))$ scaling
with $N$ \cite{Entanglement} between the two states $|0\rangle$ and
$|1\rangle$ after a time $\tau$. Therefore, in the external beam
splitter setup where these two states are spatially separated they can
be viewed as two arms of a Heisenberg limited interferometer
collectively enhanced by a factor $N$. On the other hand, in the
antiferromagnetic case, i.e. for a repulsive interaction $W>0$, the
two degenerate ground states at $J^{x}=0$
\begin{equation}
|0\rangle =|\downarrow \uparrow \cdots \downarrow \uparrow
\rangle^{z},\quad  |1\rangle =|\uparrow \downarrow \cdots \uparrow
\downarrow \rangle ^{z},  \label{qubit}
\end{equation}
are closely related to unpaired Majorana fermions which have been
considered as candidates for storing quantum information \cite{Kitaev,
  Levitov}. These states are expected to be insensitive against
perturbations since they are separated by a gap of order $W$ from the
other states of the system and are only connected via $N$-th order
perturbation theory for \textit{homogeneous} couplings
$\mathbf{J}_{l}=\mathbf{J}$. This yields stability against spin flip
errors exponentially increasing with the number of particles in the
chain $N$ and is also reflected by the scaling of the energy of the
first excited state $\epsilon_{1}\sim (J^{x}/W)^{N}$ for $J^{x}<W$
\cite{Pfeuty}.  Furthermore, if we assume that $N$ is even the states
$|0\rangle$ and $|1\rangle$ are completely insensitive to global
fluctuations of $\mathbf{J}$ since $\sum_{l}\sigma _{l}^{z}|0\rangle
=\sum_{l}\sigma _{l}^{z}|1\rangle =0$. Then the two states $|0\rangle
$ and $|1\rangle $ constitute a decoherence free subspace
\cite{dfs,Wineland} for homogeneous perturbations and can thus be used
as qubits which store quantum information reliably.

\begin{figure}[tbp]
  \centering \includegraphics[]{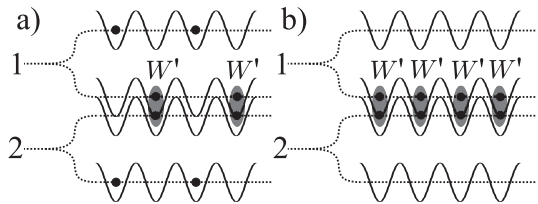}
\caption{Collectively enhanced interactions between two strings of atoms $1$
  and $2$. a) Antiferromagnetic setup: $N/2$ particles of each chain
  interact
with strength $W'$ only if they are in states $|0\rangle_1|1\rangle_2$ or $%
|1\rangle_1|0\rangle_2$ yielding a phase gate between the two qubits
implemented by those chains. b) Ferromagnetic setup: Entanglement
creation
between two chains of atoms via interactions $W'$ in the state $%
|1\rangle_1|0\rangle_2$. Impelementations with optical lattices or
atom chips, for instance, offer the scalability of the scheme.}
\label{fig4}
\end{figure}

\begin{figure}[tbp]
  \centering \includegraphics[]{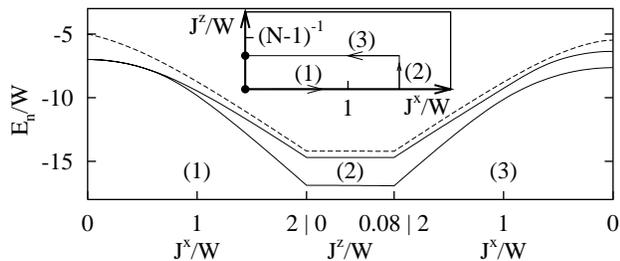} \caption{Illustration of the
    Hadamard gate for $N=8$ by adiabatically changing $J^z$ and $J^x$
    (unprotecting the quantum memory). We follow the lowest two
    eigenstates (with energies given by the solid curves) transforming
    as $|0\rangle +|1\rangle \rightarrow |0\rangle$ and $|0\rangle -
    |1\rangle \rightarrow |1\rangle$ (up to a dynamical phase) when
    changing $J^x$, $J^z$ in three steps $(1),\,(2),\,(3)$ (followed
    by turning off $J^z$ in step $(4)$) as described in the text. Note
    that if the condition $J^{z}<W/(N-1)$ is not fulfilled we get
    unwanted crossings and the first excited state after step (3) will
    not be of the form $|1\rangle$. The dashed curve shows the third
    eigenenergy of the system and the inset the path in the $J^x-J^z$
    plane.} \label{fig3}
\end{figure}

We will now discuss how to implement single and (collectively
enhanced) two qubit gates and show that our model realizes a quantum
computer with protected memory. The idea behind the two qubit phase
gate is summarized in Fig.~\ref{fig4}. Selectively overlapping the
wave functions of different two qubit states for a time $\tau_{2}$
collisional interactions of strength $W'$ between the atoms lead to an
entanglement phase $\phi _{2}=NW'\tau_{2}/2$ \cite{CCC} corresponding
to a phase gate with a truth table $|\epsilon_1\rangle
|\epsilon_2\rangle \rightarrow \exp(i \phi_2 ((\epsilon_1 +\epsilon_2)
\textrm{mod} 2)) |\epsilon_1\rangle|\epsilon_2\rangle$, with
$\epsilon_{1,2} =\{0,1\}$. Single qubit gates correspond to a general
unitary rotation of $|0\rangle$ and $|1\rangle$ (Eq.~(\ref{qubit}))
which can be decomposed in Hadamard gates $|0\rangle, |1\rangle
\rightarrow |0\rangle \pm |1\rangle$ and the creation of a relative
phase $|\epsilon_1\rangle \rightarrow \exp(i \epsilon_1 \phi_1)
|\epsilon_1\rangle$. The phase $\phi_1$ can be implemented by turning
on a trap potential creating a {\em staggered} offset of the form
$J_{l}^{z}=J^{z}(-1)^{l}$ for a time $\tau_1=\phi _{1}/2NJ^{z}$. The
idea behind the Hadamard gate is as follows: At $J^x=J^z=0$ the states
$|0\rangle$, $|1\rangle$ represent a degenerate eigenspace of $H_S$.
Turning on the field $J^x$ up to $J^x> W$, thus unprotecting the
qubit and switching it off when $J^z\ne 0$ will under appropriate
conditions induce a rotation in this space. A specific example is
illustrated in Fig.~\ref{fig3}: (1) at $J^{z}=0$ we adiabatically
switch on $J^{x}$ until $J^{x}>W$ is reached unprotecting the qubit,
then (2) we increase $J^{z}$, (3) we return adiabatically to
$J^{x}=0$, and, finally, (4) switch off $J^{z}$.

We have shown how to generate maximally entangled states of strings of
atoms in 1D pipeline configurations.  An extension of this setup
allows implementations of a quantum computing model with protected qubits.

Work supported by the Austrian Science Foundation, Deutsche
Forschungsgemeinschaft, EU Networks and Humboldt foundation.  P.\,Z.
acknowledges discussions with L.\,S. Levitov.


\begin{thebibliography}{99}
  \expandafter\ifx\csname
  natexlab\endcsname\relax\def\natexlab#1{#1}\fi
  \expandafter\ifx\csname bibnamefont\endcsname\relax
  \def\bibnamefont#1{#1}\fi \expandafter\ifx\csname
  bibfnamefont\endcsname\relax \def\bibfnamefont#1{#1}\fi
  \expandafter\ifx\csname citenamefont\endcsname\relax
  \def\citenamefont#1{#1}\fi \expandafter\ifx\csname
  url\endcsname\relax \def\url#1{\texttt{#1}}\fi
  \expandafter\ifx\csname urlprefix\endcsname\relax\def\urlprefix{URL
  }\fi \providecommand{\bibinfo}[2]{#2}
  \providecommand{\eprint}[2][]{\url{#2}}
  
\bibitem[{\citenamefont{Dumke et~al.}(2002)}]{1D}
  \bibinfo{author}{\bibfnamefont{R.}~\bibnamefont{Dumke}}
  \bibnamefont{{\em et~al.}}, \bibinfo{journal}{Phys. Rev. Lett.}
  \textbf{\bibinfo{volume}{89}}, \bibinfo{pages}{220402}
  (\bibinfo{year}{2002});
  \bibinfo{author}{\bibfnamefont{E.}~\bibnamefont{Andersson}}
  \bibnamefont{{\em et~al.}}, \bibinfo{journal}{{\em ibid.}}
  \textbf{\bibinfo{volume}{88}}, \bibinfo{pages}{100401}
  (\bibinfo{year}{2002}); \bibinfo{author}{\bibfnamefont{N.~H.}
    \bibnamefont{Dekker}} \bibnamefont{{\em et~al.}},
  \bibinfo{journal}{{\em ibid.}}  \textbf{\bibinfo{volume}{84}},
  \bibinfo{pages}{1124} (\bibinfo{year}{2000});
  \bibinfo{author}{\bibfnamefont{W.}~\bibnamefont{H{\"a}nsel}}
  \bibnamefont{{\em et~al.}}, \bibinfo{journal}{{\em ibid.}}
  \textbf{\bibinfo{volume}{86}}, \bibinfo{pages}{608}
  (\bibinfo{year}{2001});
  \bibinfo{author}{\bibfnamefont{H.}~\bibnamefont{Ott}}
  \bibnamefont{{\em et~al.}}, \bibinfo{journal}{{\em ibid.}}
  \textbf{\bibinfo{volume}{87}}, \bibinfo{pages}{230401}
  (\bibinfo{year}{2001});
  \bibinfo{author}{\bibfnamefont{N.}~\bibnamefont{Schlosser}}
  \bibnamefont{{\em et~al.}}, \bibinfo{journal}{Nature}
  \textbf{\bibinfo{volume}{411}}, \bibinfo{pages}{1024}
  (\bibinfo{year}{2001}); \bibinfo{author}{\bibfnamefont{B.~K.~Teo and
      G.~Raithel}}, \bibinfo{journal}{Phys. Rev. A}
  \textbf{\bibinfo{volume}{65}}, \bibinfo{pages}{051401}
  (\bibinfo{year}{2002}).
  
\bibitem[{\citenamefont{Greiner et~al.}(2002)}]{MottExp}
  \bibinfo{author}{\bibfnamefont{M.}~\bibnamefont{Greiner}}
  \bibnamefont{{\em et~al.}}, \bibinfo{journal}{Nature}
  \textbf{\bibinfo{volume}{415}}, \bibinfo{pages}{39}
  (\bibinfo{year}{2002}).
  
\bibitem[{\citenamefont{Jaksch et~al.}(1999)}]{CCC}
  \bibinfo{author}{\bibfnamefont{D.}~\bibnamefont{Jaksch}}
  \bibnamefont{{\em et~al.}}, \bibinfo{journal}{Phys. Rev. Lett.}
  \textbf{\bibinfo{volume}{82}}, \bibinfo{pages}{1975}
  (\bibinfo{year}{1999}).
  
\bibitem[{\citenamefont{S{\o}rensen and M{\o}lmer}(1999)}]{Molmer}
  \bibinfo{author}{\bibfnamefont{A.}~\bibnamefont{S{\o}rensen}}
  \bibnamefont{and}
  \bibinfo{author}{\bibfnamefont{K.}~\bibnamefont{M{\o}lmer}},
  \bibinfo{journal}{Phys. Rev. Lett.} \textbf{\bibinfo{volume}{83}},
  \bibinfo{pages}{2274} (\bibinfo{year}{1999});
  \bibinfo{author}{\bibfnamefont{G.~K.} \bibnamefont{Brennen}},
  \bibinfo{journal}{Phys. Rev. A} \textbf{\bibinfo{volume}{65}},
  \bibinfo{pages}{022313} (\bibinfo{year}{2002}).
  \bibinfo{author}{\bibfnamefont{E.}~\bibnamefont{Jan{\'e}}}
  \bibnamefont{{\em et~al.}}  (\bibinfo{year}{2002}),
  \eprint{quant-ph/0207011}.

  
\bibitem[{\citenamefont{Mandel et~al.}(2002)}]{Mandel}
  \bibinfo{author}{\bibfnamefont{O.} \bibnamefont{Mandel}}
  \bibnamefont{{\em et~al.}} (\bibinfo{year}{2002}),
  \eprint{cond-mat/0301169}.

  
\bibitem[{\citenamefont{Jaksch et~al.}(2000)}]{Rydberg}
  \bibinfo{author}{\bibfnamefont{D.}~\bibnamefont{Jaksch}}
  \bibnamefont{{\em et~al.}}, \bibinfo{journal}{Phys. Rev. Lett.}
  \textbf{\bibinfo{volume}{85}}, \bibinfo{pages}{2208}
  (\bibinfo{year}{2000}); \bibinfo{author}{\bibfnamefont{M.~D.}
    \bibnamefont{Lukin}}, \bibinfo{journal}{{\em ibid.}}
  \textbf{\bibinfo{volume}{87}}, \bibinfo{pages}{037901}
  (\bibinfo{year}{2001}).
  
\bibitem[{\citenamefont{DeMille}(2002)}]{DeMille}
  \bibinfo{author}{\bibfnamefont{D.}~\bibnamefont{DeMille}},
  \bibinfo{journal}{Phys. Rev. Lett.} \textbf{\bibinfo{volume}{88}},
  \bibinfo{pages}{067901} (\bibinfo{year}{2002}).
  
\bibitem[{\citenamefont{Sachdev}(2001)}]{Sachdev}
  \bibinfo{author}{\bibfnamefont{S.}~\bibnamefont{Sachdev}},
  \emph{\bibinfo{title}{Quantum Phase Transitions}}
  (\bibinfo{publisher}{Cambridge University Press},
  \bibinfo{year}{2001}).
  
\bibitem[{\citenamefont{Levitov {\em et~al.}}(2001)}]{Levitov}
  \bibinfo{author}{\bibfnamefont{L.~S.} \bibnamefont{Levitov}}
  \bibnamefont{{\em et~al.}} (\bibinfo{year}{2001}),
  \eprint{cond-mat/0108266}.
  
\bibitem[{\citenamefont{Kitaev}(2000)}]{Kitaev}
  \bibinfo{author}{\bibfnamefont{A.~Y.} \bibnamefont{Kitaev}}
  (\bibinfo{year}{2000}), \eprint{cond-mat/0010440}.
  
\bibitem[{\citenamefont{Remark}(2003)}]{Remark} \bibinfo{note}{ We
    emphasize the difference in the coupling of the Kitaev model
    \cite{Kitaev} and the spin model \cite{Levitov} to the environment
    since they are connected by a non-local transformation}.

  
\bibitem[{\citenamefont{Jaksch et~al.}(1998)}]{MottIns}
  \bibinfo{author}{\bibfnamefont{D.}~\bibnamefont{Jaksch}}
  \bibnamefont{{\em et~al.}}, \bibinfo{journal}{Phys. Rev. Lett.}
  \textbf{\bibinfo{volume}{81}}, \bibinfo{pages}{3108}
  (\bibinfo{year}{1998}).

  
\bibitem[{\citenamefont{Duan et~al.}(2002)}]{Duan} \bibinfo{note}{For
    further relizations of spin models with optical lattices see}
  \bibinfo{author}{\bibfnamefont{L.~M.} \bibnamefont{Duan}}
  \bibnamefont{{\em et~al.}}  (\bibinfo{year}{2002}),
  \eprint{cond-mat/0210564}.

  
\bibitem[{\citenamefont{Remark2}(2003)}]{Remark2}
  \bibinfo{note}{Without loss of generality we can set $J_l^y=0$ which
    can always be achieved by an appropriate rotation of
    $\mathbf{\sigma}_l$}.

  
\bibitem[{\citenamefont{Pfeuty}(1970)}]{Pfeuty}
  \bibinfo{author}{\bibfnamefont{P.}~\bibnamefont{Pfeuty}},
  \bibinfo{journal}{Ann. Phys. (NY)} \textbf{\bibinfo{volume}{57}},
  \bibinfo{pages}{79} (\bibinfo{year}{1970}).
  
\bibitem[{\citenamefont{Sackett et~al.}(2000)}]{Entanglement}
  \bibinfo{author}{\bibfnamefont{C.~A.} \bibnamefont{Sackett}}
  \bibnamefont{{\em et~al.}}, \bibinfo{journal}{Nature}
  \textbf{\bibinfo{volume}{404}}, \bibinfo{pages}{256}
  (\bibinfo{year}{2000});
  \bibinfo{author}{\bibfnamefont{M.}~\bibnamefont{Brune}}
  \bibnamefont{{\em et~al.}}, \bibinfo{journal}{Phys. Rev. Lett.}
  \textbf{\bibinfo{volume}{77}}, \bibinfo{pages}{4887}
  (\bibinfo{year}{1996}); \bibinfo{author}{\bibfnamefont{C.~J.}
    \bibnamefont{Myatt}} \bibnamefont{{\em et~al.}},
  \bibinfo{journal}{Nature} \textbf{\bibinfo{volume}{403}},
  \bibinfo{pages}{269} (\bibinfo{year}{2000}).

  
\bibitem[{\citenamefont{Khodjasteh and Lidar}(2002)}]{dfs}
  \bibinfo{author}{\bibfnamefont{K.}~\bibnamefont{Khodjasteh}}
  \bibnamefont{and} \bibinfo{author}{\bibfnamefont{D.~A.}
    \bibnamefont{Lidar}}, \bibinfo{journal}{Phys. Rev. Lett.}
  \textbf{\bibinfo{volume}{89}}, \bibinfo{pages}{197904}
  (\bibinfo{year}{2002}).
  
\bibitem[{\citenamefont{Wineland}(1992)}]{Wineland}
  \bibinfo{author}{\bibfnamefont{D.}~\bibnamefont{Kielpinski}}
  \bibnamefont{{\em et~al.}}, \bibinfo{journal}{Science}
  \textbf{\bibinfo{volume}{291}}, \bibinfo{pages}{101}
  (\bibinfo{year}{2001}); \bibinfo{author}{\bibfnamefont{D.~J.}
    \bibnamefont{Wineland}} \bibnamefont{{\em et~al.}},
  \bibinfo{journal}{Phys. Rev. A} \textbf{\bibinfo{volume}{46}},
  \bibinfo{pages}{R6797} (\bibinfo{year}{1992}).

\end{thebibliography}
\end{document}